\documentclass{article}
\usepackage{arxiv}

\usepackage[utf8]{inputenc}
\usepackage[T1]{fontenc}
\usepackage{amsmath}
\usepackage{amsfonts}
\usepackage{amssymb}
\usepackage{booktabs}
\usepackage{nicefrac}
\usepackage{microtype}
\usepackage{graphicx}
\usepackage[most]{tcolorbox}
\usepackage[numbers]{natbib}
\usepackage{hyperref}
\usepackage{cleveref}
\usepackage{needspace}
\usepackage{enumitem}

\hypersetup{
  colorlinks=true,
  linkcolor=blue,
  citecolor=blue,
  urlcolor=blue,
  pdftitle={Safety Features for a Centralized AGI Project},
  pdfauthor={Sarah Hastings-Woodhouse}
}

\usepackage{titlesec}

\titleformat{\section}
  {\normalfont\Large\bfseries}{\thesection}{1em}{}[\nopagebreak]
  
\titleformat{\subsection}
  {\normalfont\large\bfseries}{\thesubsection}{1em}{}[\nopagebreak]
  
\titleformat{\subsubsection}
  {\normalfont\normalsize\bfseries}{\thesubsubsection}{1em}{}[\nopagebreak]

\usepackage{mdframed}
\usepackage{tikz}
\usepackage{xcolor}  % for hex colors

\definecolor{recommendgreen}{HTML}{e2fac2}  % Light green

\newenvironment{keyrecommendations}[1][]{%
  \ifstrempty{#1}%
  {\mdfsetup{%
    frametitle={%
      \tikz[baseline=(current bounding box.east),outer sep=0pt]
      \node[anchor=east,rectangle,fill=recommendgreen]
      {\strut Key recommendations};}}
  }%
  {\mdfsetup{%
    frametitle={%
      \tikz[baseline=(current bounding box.east),outer sep=0pt]
      \node[anchor=east,rectangle,fill=recommendgreen]
      {\strut Key recommendations:~#1};}}%
  }%
  \mdfsetup{innertopmargin=10pt,linecolor=recommendgreen,%
    linewidth=2pt,topline=true,
    frametitleaboveskip=\dimexpr-\ht\strutbox\relax,
    nobreak=true}  % <-- This prevents page breaks
  \begin{mdframed}[]\relax%
}{\end{mdframed}}

\title{Safety Features for a Centralized AGI Project}
\author{
  Sarah Hastings-Woodhouse \\
  Pivotal Fellowship \\
  \texttt{sarah.hastings.woodhouse@gmail.com}
}
\date{\today}

\begin{document}
\maketitle

\begin{abstract}
Recent AI progress has outpaced expectations, with some experts now predicting AI that matches or exceeds human capabilities in all cognitive areas (AGI) could emerge this decade, potentially posing grave national and global security threats. AI development is currently occurring primarily in the private sector with minimal oversight. This report analyzes a scenario where the US government centralizes AGI development under its direct control, and identifies four high-level priorities and seven safety features to reduce risks.
\end{abstract}

\keywords{Artificial General Intelligence \and AI Safety \and AI Governance \and AI Policy \and National Security}

\section{Executive Summary}

\subsection{Context}

Recent AI progress has outpaced expectations, with some experts now predicting AI that matches or exceeds human capabilities in all cognitive areas (AGI) could emerge this decade, potentially posing grave national and global security threats. AI development is currently occurring primarily in the private sector with minimal oversight. This report analyzes a scenario where the US government centralizes AGI development under its direct control, and identifies four high-level priorities and seven safety features to reduce risks.

\subsection{High level priorities for a government-led project}

\begin{enumerate}
\item \textbf{Assess Alignment Difficulty}: Continuously evaluate whether the project is on track to solve the problem of controlling systems more intelligent than humans

\item \textbf{Maintain Optionality}: Retain the flexibility to pivot away from rapid AGI development if evidence suggests it cannot be done safely

\item \textbf{Envision the End-State}: Plan for post-AGI scenarios from the beginning
\end{enumerate}

\subsection{Safety Features}

\textit{1. Information Escalation}

Establish robust reporting requirements triggered by regular intervals, compute scaling thresholds, and detection of concerning capabilities. Create protected communication channels including emergency protocols, formal dissent mechanisms, and enhanced whistleblower protections.

\textit{2. Pause Protocols}

Define clear risk thresholds beyond which development must halt. Enable emergency pauses through both bottom-up (technical staff) and top-down (leadership) channels. Establish formal checkpoints requiring affirmative safety cases before proceeding to next development stages.

\textit{3. Board Oversight}

Establish a board with binding authority, composed of technical experts qualified to assess AI risk. Grant this board legal authority through Congressional action to approve or disapprove key project decisions, particularly regarding training new models or expanding system permissions.

\textit{4. Audit and Risk Monitoring}

Implement specialized internal audit mechanisms beyond standard government oversight to scrutinize risk management practices. Create a dedicated risk monitoring team separate from those managing risks to quantify catastrophic risk probabilities at each development checkpoint.

\textit{5. Geopolitics Division}

Form a specialized body to track adversaries' AI progress and prepare for high-stakes diplomatic scenarios. Develop detailed response plans for critical situations including: discovery of unacceptable risks, threats of military action, adversary AGI development, model weight theft, and internal power grabs.

\textit{6. Designated Verification Project}

Develop technologies to verify compliance with potential international agreements limiting AI development. Explore hardware-enabled mechanisms that embed regulatory controls directly into AI chips and pursue international collaboration on verification standards.

\textit{7. Automated Research Plan}

Establish clear boundaries on AI-enabled research acceleration, with Board-approved red lines on system authority and explicit oversight requirements that scale with acceleration capabilities. Develop comprehensive research agendas for using automated systems to enhance safety rather than just capabilities.

\section{Introduction}

\textbf{Over the past few years, rapidly accelerating AI progress has surpassed expectations} -- with many industry insiders now predicting AI that is competitive with humans in every cognitive domain, sometimes referred to as Artificial General Intelligence (AGI), before the end of the decade.

Frontier AI systems already exhibit extremely impressive capabilities, with OpenAI's latest reasoning model, o3, \cite{openai_openai_2024} outcompeting PhD-level human experts at advanced science questions, outranking more than 99\% of human competitive coders on the company's CodeForces benchmark, and scoring 25\% on FrontierMath, the world's toughest mathematics benchmark (its predecessor, o1, released just months earlier, scored 2\%).

These trends are expected to continue. Forecasts suggest that, even taking into account constraints on energy, data and chip manufacturing, it will be possible to train models with 10,000x the compute used for OpenAI's GPT-4 (arguably the most capable family of models for most of 2023) by 2030\cite{sevilla_can_2024}.

\textbf{Experts have warned that the arrival of very powerful AI could pose grave national and global security threats}, up to and including human extinction \cite{noauthor_statement_nodate}. Risks could emerge from misuse of AI models, for example to assist in the development of novel biological weapons, from loss of control over powerful, goal-directed systems, or from disturbances to the international balance of power that trigger military escalation.

\textbf{The development of powerful AI is currently taking place in the private sector, with very little government oversight}. Executive Order 14110 \cite{noauthor_safe_2023}, which required companies training large models to share the results of safety testing with the government -- the only binding regulation specifically targeting frontier AI development on the federal level -- was recently overturned. \cite{noauthor_removing_2025}.

This lack of state involvement in AI development marks a departure from historical technologies that shared its transformative potential and strategic significance, such as nuclear weapons, space exploration and telecommunications. Ensuring the safety and security of very powerful AIs may exceed the capabilities of the private sector. A report by the RAND corporation found that preventing the theft of model weights by motivated national states or other well-resourced actors may be impossible without government assistance\cite{nevo_securing_2024}.

As the US government becomes more aware of AI's national security implications, they may seek greater control over its development. The US-China Economic and Security Review Commission's 2024 report, for example, urges a ``Manhattan Project-like program dedicated to racing to and acquiring an Artificial General Intelligence (AGI) capability''\cite{noauthor_2024_2024}.

\textbf{The government may exert control over frontier AI development and deployment to various extents, using a variety of mechanisms\cite{cheng_soft_nodate}.} This report focuses on a scenario in which it has taken action on the extreme end of this spectrum, by centralizing AGI development into a single project under its direct control.\footnote{This may be possible using authorities in the Defense Production Act, through which the Executive can compel private companies to accept contracts that are ``necessary and appropriate to promote the national defense''. The Act also grants the President the authority to ``control the distribution of any supplies of materials, services, and facilities in the marketplace'' that are essential to this effort. This may enable a de-facto ban on frontier training runs outside of a government project due to monopolization of the required computing resources. This would not require literal nationalization of frontier AI companies, though it could produce a similar result in practice. A literal reading of the Defense Production Act appears to allow for such measures, though in practice, their feasibility is contested -- especially without congressional and/ or judicial approval.} Experts disagree as to the likelihood of this scenario -- but if it is plausible, it warrants analysis.

\textbf{If the design and execution of a centralized AGI project is insufficiently cautious and secure, it may do more damage to national security than the status quo}. This report describes some measures that could be taken to promote the success of such a project and reduce its risks. It describes seven ``safety features'', informed by semi-structured interviews with experts in AI safety and governance. It does not take a position on the likelihood or merit of centralization, but simply considers how to mitigate potential downsides should it occur.

\section{High-level priorities}

This section proposes \textbf{four guiding principles} that should inform the design and execution of a centralized AGI project. They rest on the assumption that AGI is likely to pose extreme risks that threaten global security if not developed cautiously.

\subsection{Assess the difficulty of alignment}

A central priority of the project should be assessing the difficulty of aligning powerful AI systems. This is a question on which there is little scientific consensus. Anthropic acknowledges that ``[developing] advanced AI systems that are broadly safe and pose little risk to humans...could lie anywhere on the spectrum from very easy to impossible''\cite{noauthor_core_nodate}. Figure \ref{fig:how-hard} illustrates this view:

\begin{figure}
    \centering
    \includegraphics[width=0.8\linewidth]{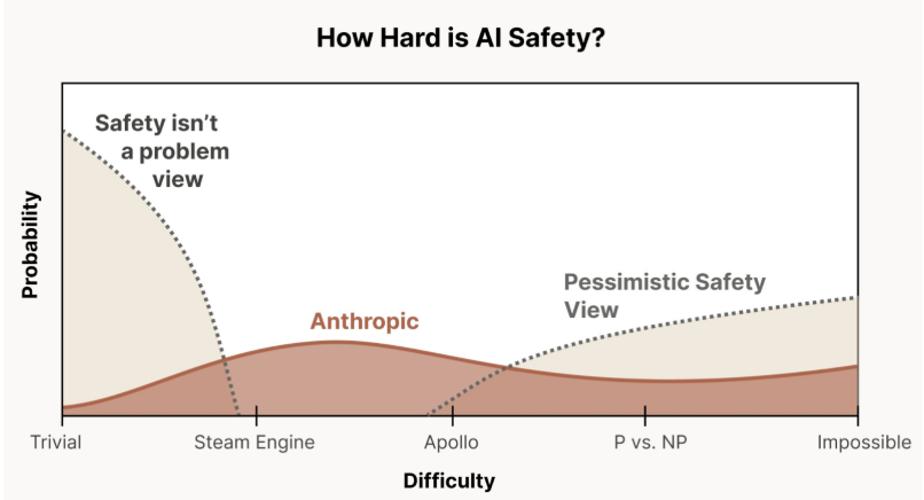}
    \caption{How hard is AI safety? \cite{olah_how_2023}}
    \label{fig:how-hard}
\end{figure}

We may not be able to take an iterative approach to solving this problem, since loss of control over AIs more powerful than ourselves would be irreversible and likely catastrophic. Therefore, the project should invest significant time and resources into gathering evidence about how hard it will be to align smarter-than-human AIs, and forecasting whether it is on track to achieve this before their arrival.

For example, difficulty in addressing phenomena such as reward-hacking \cite{skalse_defining_2025}, deception \cite{park_ai_2023} and vulnerability to jailbreaks \cite{zhou_easyjailbreak:_2024} may be evidence for alignment being more challenging, while methods for reducing them that reliably scale with model size may be evidence for its relative ease. The project should establish clear metrics for tracking progress on these indicators and reassess the likelihood of successful alignment as capabilities advance. If the project can recognize preemptively that it is not on track to align powerful AI, it may be able to halt development before catastrophe occurs.

\subsection{Maintain optionality}

If the US pursues a government-led AGI project, it is likely to be with the ultimate goal of attaining AGI capabilities before its adversaries, bolstering national security and preventing AI-enabled power grabs by authoritarian regimes. Stopping autocratic states from leveraging powerful AI to achieve global dominance is essential, and \textbf{should} be a key priority of the United States. However, the leadership of a government-led project should remain open to the possibility that racing to develop AGI as fast as possible may not be a viable means of accomplishing this goal.

Ensuring that AI systems exceeding human intelligence behave as intended, or are ``aligned'', is an open research problem. It may not be solvable within the next few years -- the same timeframe in which many experts now predict such advanced systems will be developed. Powerful and uncontrolled AI could pose extreme risks to humanity, regardless of whether it is built in the US or elsewhere. Therefore, project leadership should \textbf{retain the option} of abandoning the goal of quickly building AGI, should they receive compelling evidence that doing so safely is not feasible. As discussed in sections 2 and 6, this can be achieved through implementing the protocols needed to pause the project in an emergency and planning for high-stakes diplomacy with adversaries.

\subsection{Envision the end-state of the project}

Ideally, planning for post-AGI ``end-states'' should begin as soon as the project itself is underway, if not long before. A plan that starts and ends with crossing the AGI finish-line first is not sufficient. Leadership should consider how they will manage threats of military action or attempts at sabotage by hostile states \cite{noauthor_deterrence_nodate} as the project approaches its final stages. They should make a detailed plan for how they will execute on a decisive advantage provided by AGI capability that is non-violent, allows adversaries to retain their sovereignty and does not provoke retaliatory action. This will be a very fine needle to thread, and may prove impossible -- in which case the project should be willing to deprioritize this goal and in favor of diplomatic efforts.

\section{Organizational structure and information flows}

\begin{figure}
    \centering
    \includegraphics[width=1\linewidth]{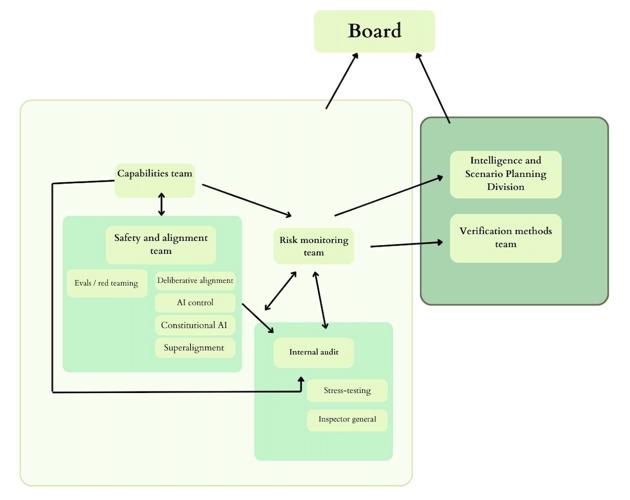}
    \caption{Organizational chart}
    \label{fig:info-flow}
\end{figure}

Figure \ref{fig:info-flow} shows the basic organizational structure of a government-led AGI project. It is intended to illustrate the information flows between each of the bodies proposed within the safety features, not to be a comprehensive taxonomy of personnel that might be involved in such a project. The arrows illustrate information flows between bodies.

The Safety and Alignment team contains a number of “sub-teams” dedicated to various safety research agendas such as \href{https://openai.com/index/introducing-superalignment/}{Superalignment} and \href{https://www.anthropic.com/research/constitutional-ai-harmlessness-from-ai-feedback}{Constitutional AI}. The examples listed are intended to be illustrative and are not recommendations or endorsements. 

We recommend that the project allocate an individual team to each agenda it decides to pursue in order to ensure diversification of research directions – as it is not currently clear which, if any, will yield results. Sub-teams should meet regularly to discuss results, and could merge to pursue a single direction if it appears particularly promising. 

\section{Safety features}

This section proposes seven ``safety features'' to reduce the global and national security risks of a government-led AGI project. Note that the features identified vary in their level of interdependence. Some could likely be implemented in isolation, while others would require coordinated implementation to be effective.

\subsection{Information escalation}

\begin{keyrecommendations}
\begin{enumerate}
\item
  Reporting requirements (from technical teams to project leadership),
  designed to detect dangerous capabilities. These requirements should
  specify the precise capabilities that need to be reported and the
  evaluations needed to detect them.
\item
  Information channels suitable for emergency communication about
  imminent AI-related threats. This could be modelled on existing
  channels for rapid communication of intelligence such as the
  The Critical Information Communications (CRITIC) system.
\item
  A dissent channel for members of technical teams to report concerns
  about safety risks. This should be a \textbf{resolution} (rather than
  just a reporting) mechanism. The dissenting party should be offered a
  surrogate to represent them, to preserve anonymity.
\item
  Augment these information channels with whistleblower protections to
  ensure individuals are not deterred from coming forward. Consider
  protecting dissent channels under the Whistleblower Protection Act
  (2012), and/ or broadening the ``substantial and specific danger''
  requirement in the National Defense Authorization Act for Fiscal Year
  2013 to encompass ``reasonable belief'' that disclosure would reduce
  the probability of a catastrophic incident.
\end{enumerate}
\end{keyrecommendations}

A key priority of the project should be ensuring that its leadership receives timely notice of imminent and emerging risks. If these risks arise, they will first be known to technical teams working directly with models. This will need to be efficiently escalated up to project leadership, and to the risk monitoring and internal compliance teams (see Figure 1).

Achieving this will require \textbf{robust reporting requirements} and \textbf{protected information channels}.

\subsubsection{Reporting requirements}

There are a number of reasons why the development of AI models might necessitate thorough model testing and risk reporting much earlier in the development cycle than for other technologies.

Firstly, catastrophic risks from powerful AI models could arise during model training or internal deployment \cite{stix_ai_2025}. For example, the weights of an internally deployed model could be stolen by a malicious actor, or the model could be misused by its developers. Even before training is complete, a sufficiently powerful model could escape human control by exfiltrating its weights from company-controlled hardware, accelerating AI R\&D \cite{noauthor_evaluating_2024} to a degree that humans can no longer reliably monitor, or manipulating its human overseers \cite{noauthor_measuring_nodate} into taking actions that undermine their interest.

Secondly, AI progress is also unpredictable. ``Emergent capabilities''  can come as a surprise even to model developers\cite{wei_emergent_2022}, and the relationship between training compute and performance \cite{heim_training_2024} is not well understood\cite{heim_training_2024}.

Finally, future improvements in AI capabilities may be discontinuous \cite{noauthor_will_nodate}, meaning that risks could emerge with little warning.

\textbf{For these reasons, we recommend that a government-led project impose reporting requirements on technical teams:}

\begin{enumerate}
\item \textbf{at a regular cadence (eg quarterly)}
\item \textbf{every time effective compute is scaled up by a certain amount}\footnote{Existing lab safety frameworks provide useful precedent here. OpenAI's Preparedness Framework sets the most conservative threshold for running evaluations and reporting their results, at every 2x in effective compute (or every time there is a new algorithmic breakthrough). Anthropic and Google DeepMind set this threshold at 4x and 6x respectively. ``Effective compute'' is a better proxy for capability than pre-training compute alone, because it takes into account other drivers of progress such as algorithmic efficiency. However, the project should prepared to regularly review and update these thresholds \cite{noauthor_response_nodate} to account for new breakthroughs.}
\item \textbf{whenever a model exhibits one of a pre-specified list of capabilities.}
\end{enumerate}

Combining these three reporting mechanisms would increase the probability of discovering imminent threats before they materialize.

\textbf{These requirements should clearly define the exact capabilities that need to be reported, and the evaluations which should be performed to detect them}. There are several regulatory precedents and industry standards that could prove instructive, though none meet this bar.

Prior to its reversal in January 2025, President Biden's 2023 Executive Order \cite{noauthor_safe_2023} ``Safe, Secure, and Trustworthy Development and Use of Artificial Intelligence'' established reporting protocols for organizations developing frontier models trained above a computational threshold of 10\^{}26 floating-point operations (FLOP). Companies were obligated to notify the government about plans to train such frontier models, outline plans to mitigate potential hazards, and disclose findings from red-team exercises designed to identify critical risks, including biological and cybersecurity threats as well as capacity for self-replication or autonomous propagation. The order was later supplemented by NIST's AI Risk Management Framework\cite{noauthor_ai_2021}, which offers broad guidance on safety-testing practices, and recommends that companies determine their own ``risk tolerances''. But even while the Executive Order was in effect, companies were never required to perform specific evaluations or to report the emergence of particular capabilities.

Another source we might consult is the safety frameworks \cite{noauthor_common_2025} released by each of the frontier AI companies. These generally establish ``tripwire capabilities'' which trigger certain responses -- such as disclosure to internal governance bodies or government authorities, the implementation of more stringent risk mitigations or even halts to deployment and development (as we'll discuss in the next section). However, these thresholds are not always well-defined. For example, Anthropic has yet to specify the capabilities that would classify a model as AI Safety Level 4 (ASL-4)\cite{noauthor_announcing_nodate}, even as they expect to reach this level within the next two years. xAI's Risk Management Framework \cite{noauthor_xai_2025} is, so far, the only document to specify the benchmarks and evaluations it will use to track dangerous capabilities, though the company has not determined what specific scores in these tests will be sufficient to trigger particular responses.

Work done outside of either labs themselves or government legislation may also be useful in forming reporting requirements for a government-led project. Some work has been done on ``intolerable risk thresholds'', which could mark the point at which models begin to pose unacceptable risks. For example, a report from UC Berkeley's Center for Long-Term Cybersecurity provides specific threshold recommendations and detailed case studies across eight risk categories including CBRN misuse, model autonomy and deception -- though it does not claim to capture every intolerable risk in its taxonomy\cite{raman_intolerable_nodate}.

\textbf{The thresholds required to \textit{report} a capability within the context of a government-led project should be lower than those at which risks become unacceptable.} \textbf{The project should ensure there is a significant buffer between capabilities that must be reported and those that would trigger a pause in development or deployment}.

In the context of a government-led project aiming to build extremely powerful and potentially dangerous AI systems, a better-specified set of reporting requirements and risk thresholds than is provided by the above frameworks will be essential. Since the science of evaluating AI models is still nascent, a significant research effort will be needed to achieve this. This should be among the early priorities of the project. Thresholds for reporting should be set very conservatively while this effort is underway.

\subsubsection{Information channels}

Information about potentially dangerous model capabilities should be reported to:

\begin{enumerate}
\item Project leadership (the Board and nominated personnel within the US Executive Branch)
\item The internal audit team
\item The risk monitoring team
\end{enumerate}

\textbf{Emergency communication}

The reporting mechanisms described above could be triggered both expectedly and unexpectedly. Reports that are due on a predetermined basis, or each time effective compute is scaled up by a certain amount, can take a similar form to the model scorecards currently produced by AI companies, and simply disseminated among the relevant teams. However, reporting of emergent capabilities could have to happen on very short notice, and may therefore require specific mechanisms to ensure timely escalation.

One useful precedent for time-sensitive reporting of imminent risks is \textbf The Critical Information Communications (CRITIC) system \cite{noauthor_5_nodate}, which was established in 1959 and is used to transmit information about urgent national security threats from US government agencies to top officials, including the President. A declassified report from the CIA Historical Review Program reveals that it was quite successful at increasing the speed at which critical intelligence was communicated to the White House, which was reduced from nine and a half hours to less than one hour in its first year of operation\cite{tidewell_notes_nodate}. Public information about how CRITIC currently operates is limited, except that cables are sent via the State Messaging and Archive Retrieval Toolset (SMART) Center over secure intelligence network ClassNet, and that they automatically take precedence over others.

The most detailed overview \cite{noauthor_handling_1979} of CRITIC operation is from 1979 and may no longer be accurate, but can still help inform a similar approach in the context of a government-led AGI project. The National Security Agency (NSA) was responsible for operating an automated system to simultaneously deliver cables to all intended recipients, while overriding less urgent transmissions. Any government official could author a cable, but the NSA had sole authority to transmit it to its final destination.

An adapted CRITIC system for reporting dangerous AI capabilities could take a similar form (though authority to author one would have to be expanded to include government contractors). The NSA, with its expertise in managing classified communications infrastructure, could be an appropriate body for overseeing the transmission of cables. It could operate a similar automated system for simultaneously delivering them to both the President and relevant teams within the project.

The State Department also conducts regular CRITIC Exercises, which aim to simulate the delivery of real cables in under ten minutes. This kind of exercise could form the basis of emergency drills to help prepare for the detection of dangerous AI capabilities.

\textbf{Dissent}

As well as imminent threats from dangerous capabilities, members of technical teams may become aware of emerging risks or dangerous practices. The project should implement measures allowing them to report these concerns without fear of retaliation.

These could be modelled on the ``dissent channels'' which exist at several government agencies to allow employees to express disagreement with government policies or propose alternatives, such as the State Department \cite{noauthor_2_nodate}, the Nuclear Regulatory Commission \cite{noauthor_nrc_2015} (NRC) and the Department of Energy \cite{noauthor_doe_nodate}. These channels can vary considerably in their scope and implementation. Some are available only to direct government employees, while others can also be used by contractors -- a dissent channel set up in the context of a government AGI project would obviously need to be available to both.

Another important difference is between \textbf{reporting} and \textbf{resolution} mechanisms; the former exists primarily to ensure leaders are aware of dissenting opinions and places little obligation on them to respond, while the latter is a more involved process that aims to achieve consensus on a contested issue. A government-led AGI project should aim to establish a resolution mechanism. Generally speaking, resolution mechanisms are used in agencies where dissents are technical in nature. This lends itself nicely to the AGI context, where the most consequential dissents are likely to concern unsafe model development.

The most instructive example may be the Nuclear Regulatory Commission's Differing Professional Opinions Process \cite{noauthor_doe_nodate}, where employees might report, for example, potential safety failures in nuclear power plant operations. Cases are reviewed by a three-person panel, ideally formed of people outside of the submitter's direct chain of command and with technical expertise in the issue at hand.

Members of the \textbf{internal audit team} (see section 4) may be ideal candidates for an equivalent process within the AGI project. The dissenting party can also remain anonymous through use of a \textbf{surrogate} who represents them during proceedings. This could prove an extremely useful practice, since fear of retaliation has often deterred use of existing dissent channels (and retaliation has been occasionally known to occur, despite most agencies formally prohibiting it). The submitter can also appeal the final decision of the panel if they do not feel their concerns have been adequately addressed.

\textbf{Whistleblower protections}

The information channels described above will need to be augmented with whistleblower protections to ensure that concerned individuals are not deterred from coming forward.

The extent to which existing legislation would protect whistleblowers within the context of a government-led AGI project is unclear. According to the government's Office of Special Counsel, whether statutory protections cover the use of dissent channels is currently ``untested'' \cite{schooten_stifling_nodate}, meaning any such mechanism set up within an AGI project would face similar ambiguity. The National Defense Authorization Act for Fiscal Year 2013 (NDAA 2013) instituted a pilot programme for extending whistleblower protections to government contractors, which was made permanent in 2016. It covers contractors reporting ``a substantial and specific danger to public health or safety'' -- which is promising in the AGI context. However, the nature and timing of risks posed by powerful AI may be extremely hard to determine in advance \cite{field_why_2025}, and various experts offer disparate assessments of risk \cite{field_why_2025} -- which may make the presence of a ``substantial and specific danger'' difficult to prove.

Legislative changes could help address these gaps. A report \cite{schooten_stifling_nodate} into dissent channels by Project on Government Oversight recommends that Congress amend the Whistleblower Protection Act (2012) to specify their inclusion. This specification would also need to be made in the NDAA 2013 to ensure protection for contractors working in a government-led AGI project. One proposal for an AI whistleblowing regime suggests broadening the ``substantial and specific danger'' requirement to encompass ``reasonable belief'' that disclosure would reduce the probability of a catastrophic incident\cite{wu_ai_2025}. This would ensure that AI dangers of uncertain likelihood can be safely reported.

\subsection{Emergency protocols}

\begin{keyrecommendations}
\begin{enumerate}
\item
  Develop ``limit evals'' designed to detect capabilities that could
  pose extreme risks.
\item
  Establish clear intolerable risk thresholds or ``tripwire
  capabilities'' beyond which deployment or development must be halted
  until effective mitigations are put in place.
\item
  Appoint a point of contact who is the sole intermediary required to
  execute a pause instruction should a member of technical staff request
  one.
\item
  Establish both ``bottom-up'' and ``top-down'' pause protocols, so that
  any member of staff concerned about imminent and extreme risks can
  trigger an initial halt. Require increasingly senior sign-off for
  longer pauses.
\item
  Before resuming training, require technical teams to present a safety
  case to the board, which makes a clearly-evidenced argument for why
  doing so would not pose intolerable risks. Safety cases should include
  emergency response plans to mitigate against extreme risks that emerge
  after internal deployment.
\end{enumerate}
\end{keyrecommendations}

If evidence of imminent risk emerges, the project may need to pause AI training on very short notice. Additionally, it must have measures in place to ensure a durable pause is possible in the event that internally deploying a trained model, or training one at a new compute threshold, is determined to be unacceptably risky. This can be achieved through a combination of defining \textbf{intolerable risk thresholds}, instituting policies for \textbf{emergency shutdown of training runs}, and requiring \textbf{safety cases} for high-risk activities.

\subsubsection{Intolerable risk thresholds}

As discussed in the previous section, the project will need to establish clear intolerable risk thresholds or ``tripwire capabilities'' beyond which deployment or development must be halted until effective mitigations are put in place.

Frontier AI companies have already released safety frameworks which commit to do this in principle, but risk thresholds are not well-defined, and safety frameworks do not generally specify the evaluations that would be needed to detect them. Even if the commitments in these safety frameworks were perfectly specified, the competitive pressures driving companies to develop increasingly capable models are likely to undermine these safeguards. This is explicitly acknowledged \cite{noauthor_announcing_nodate} in Antrophic's Responsible Scaling Policy, which allows the company to lower its own required safeguards in the event that it is being outpaced by a less cautious actor.

A government-controlled AGI project would possess a key advantage -- it could standardize a single set of ``pause thresholds'' without concern that race dynamics between multiple companies will render them ineffective, or that the actor with the least robust safeguards will inevitably gain competitive advantage.

However, there remains much research to be done in determining and operationalizing these thresholds. Many existing evaluations focus on tasks far below the level that would pose catastrophic risks -- for example, the illustrative evaluations offered for CBRN risks in OpenAI's Preparedness Framework are designed to determine whether models remain in its ``low'' risk category (meaning they can ``provide information relevant to creating CBRN threats with comparable utility to existing resources'')\cite{openai_our_nodate}.

The project should invest considerable effort into developing ``limit evals'' \cite{noauthor_sketch_nodate} that would actually track capabilities sufficient to cause large-scale damage, such as being able to walk an amateur through the end-to-end process of designing and deploying a bioweapon. These are likely to be costly to run, but could be neatly tied to intolerable risk thresholds.

Several efforts have been made to establish intolerable risk thresholds from frontier AI systems\footnote{See \cite{raman_intolerable_nodate}, \cite{noauthor_risk_nodate}, \cite{noauthor_sketch_nodate}.}, though none claim to provide a complete taxonomy. Developing a comprehensive set of thresholds and accompanying evaluations may take considerable time and research. The project can \textit{begin} defining unacceptable risks and benchmarking models using simpler, lower-cost evaluations ahead of this point. Indeed, the US government can begin accelerating research in this area, for example through adequate resourcing of the US AI Safety Institute, before a centralized project is launched. However, as with the core problem of alignment, the project should be open to the possibility that developing a mature science of model evaluation may not be possible in a short timeframe. Clear evidence of this should itself trigger a pause in advancing model capabilities.

\subsubsection{Emergency pauses}

It may be possible for AI models to pose substantial risks before they have completed training\cite{noauthor_ai_2025}. The authority to pause a training run should ideally extend all the way to low-level technical staff who may be first to learn of an imminent threat, though extending this pause beyond an initial period would require higher-level sign-off. It should be possible to delegate a pause instruction down the command chain as well as to escalate one up it.

\textbf{Bottom-up escalation}

Suppose a member or members of a technical team suspects that further development of a model may produce extreme and imminent risks. These could stem from a system that autonomously pursues undesirable goals, attempts to break out of its training environment, is developing persuasive capabilities sophisticated enough to manipulate its overseers, or crosses any of the predetermined thresholds discussed above. Even if this staff member lacks seniority, they should possess the authority to immediately halt the training run.\footnote{The project might understandably be concerned about insider threats or attempts to sabotage progress towards AGI under the guise of safety concern. An alternative model could permit concerned staff members to submit pause \textit{requests} to the internal compliance and / or risk monitoring teams, which must be reviewed within a short period. Whether this will be sufficient to mitigate imminent risks, however, will depend on precisely how time-sensitive loss-of-control threats turn out to be. This is a matter of live debate -- a speculative but not thoroughly rebutted possibility is a ``hard takeoff'' in which AI self-improves to become radically superhuman within hours or even minutes. ``Points of no return'' could also occur well before takeover, and with very little notice.

The project should work to gather as much empirical evidence about the likelihood of such scenarios as it can -- but they may be very difficult to rule out the possibility in advance. This means the most risk-averse policy would grant immediate pause authority to all staff members within the project. This trades off against some risk of sabotage, though requiring senior sign-off for pauses lasting more than an hour should reduce this risk.} The technical process for pausing a single training run is uncomplicated -- though in practice, it would be elongated by the need to obtain the correct permissions.

The project should appoint a \textbf{point of contact} who is the sole intermediary required to execute a pause instruction should a member of technical staff request one. This post could be rotated among more senior members of staff on a regular basis. While undertaking this duty, the designated staff member should be insulated from interactions with the model, to avoid the possibility of being manipulated through persuasive capabilities that the model might have developed.

Once an initial halt has been instituted, the staff member(s) responsible will need to escalate this information up the command chain. Increasingly senior levels of sign-off would be required the longer the pause is extended. This could involve, for example, producing a justification report for the \textbf{internal compliance} and \textbf{risk monitoring teams} within the hour, and further escalation to the board and a designated senior member of the Executive Branch within several more if the threat is considered credible. Assuming that concerns are not resolved at a lower level\footnote{Though the project should aim to resolve safety issues at a lower level, it should be wary of hasty attempts to ``patch'' bad behavior that ultimately reinforce it. Research has shown that such attempts can lead to phenomena such as reward-hacking \cite{pan_effects_2022}, sycophancy \cite{sharma_towards_2025} and deception \cite{park_ai_2023}.}, final authority to either resume training or extend the pause should land with the Board.

Broad authority to raise and escalate concerns about safety failures, even among low-level staff, is a common and effective practice in high-stakes environments. For example, a key tenet of Crew Resource Management, which revolutionized aviation safety protocols following several fatal accidents in the 1970s, is ``assertiveness training'' for junior crewmembers. Data from both experimental simulations and real-world events shows that communication failures, particularly from the bottom up, are \href{https://d1wqtxts1xzle7.cloudfront.net/55733875/CREW_RESOURCE_MANAGEMENT-libre.pdf}{often the cause} of accidents.

Automotive manufacturer Toyota implements \textit{Jidoka} as part of its production system, which it describes as ``the principle of designing equipment to stop automatically and to detect and call attention to problems immediately whenever they occur''. Anyone working on a production line at a Toyota factory is authorized to activate an ``Andon cable'' \cite{noauthor_andon_2016} if they suspect a threat to vehicle quality or safety, which triggers an automatic stop to production while the issue is dealt with.

In high-risk work environments such as construction and manufacturing, Stop Work Authority (SWA) programs are a common feature of health and safety regulation\cite{noauthor_stop_2024}. These empower employees to voice concerns about safety risks and refuse to work in hazardous conditions without fear of retaliation.\footnote{The applicability of Stop Work Authority in the context of AGI development could be an interesting area for further research. SWA is not technically enshrined in federal law, but several past court rulings have set precedents that closely align with its principles. For example, Whirlpool Corporation v. Marshall \cite{noauthor_whirlpool_nodate} (1980) established that workers have the right to refuse work without being penalized if they reasonably fear serious injury or death under the 1970 Occupational Health and Safety Act \cite{noauthor_osh_1970}. Of course, any right granted here would be to an individual contractor to refuse to perform his/ her duties, not to trigger a project-wide halt in operations (and what constitutes a reasonable belief of danger would likely come into question in the AI context).}

A government-led AGI project should strongly consider adopting similar principles. Because anticipating risks from AI is not a settled science, there is likely to be a spectrum of opinion about the likelihood and imminence of threats within the project -- and without robust safeguards, organizational incentives will naturally favor those with less conservative safety assessments\cite{armstrong_racing_2016}. This will almost certainly result in overlooked and possibly catastrophic safety failures.

\textbf{Top-down commands}

While members of technical teams will likely be the first to become aware of risks arising from the emergence of dangerous capabilities, project leadership may be the first to register others, such as a diplomatic crisis with an adversary state who is concerned by US progress towards AGI.

Unilateral ability to order an immediate halt to training runs within the project should be granted to the President. The datacenter(s) involved in the training run should be air-gapped for security, precluding the possibility of a remote shutdown device -- so the President will need to execute the order through the designated individual within the project granted pause authority.

\subsubsection{Pause checkpoints}

Catastrophic loss-of-control could theoretically arise at any point during the development of a model, but becomes substantially likelier as the model becomes more powerful and is granted more and more permissions. Key thresholds could include the commencement of post-training, or the internal deployment of a model for various uses such as accelerating research into capabilities, safety, governance or security.

Determining if it is safe to proceed to the next stage of model development or deployment -- whether to connect a trained model to the internet or to trust it with particular critical decision-making responsibilities, for example -- is complicated by the unreliability of evaluations and the black-box nature of AI systems.

There is evidence that models can develop situational awareness \cite{laine_me_2024}; the ability to distinguish between their testing and deployment phases, and to understand the motivations of their overseers. They may therefore employ deceptive tactics such as sandbagging \cite{weij_ai_2025} (strategically underperforming during capability tests) or concealing their true goals \cite{noauthor_alignment_nodate}, which could undermine the effectiveness of pre-deployment testing. This may cause the project to internally deploy or otherwise grant permissions to a model which appears safe, but in fact has the capacity and inclination to cause catastrophic harm. Additionally, if testing proves inconclusive, the project may be incentivized to tolerate some risk of catastrophe due to fear of being overtaken by adversaries.

\textbf{To mitigate against this, the project should establish a series of checkpoints within the development of a single model, at which technical teams must present an affirmative safety case (discussed further below) to the Board. The Board must vote by a supermajority to permit progression beyond the checkpoint}. Potential checkpoints include:

\begin{enumerate}
\item \textbf{Before pre-training}: Once plans for the training run have been finalized, but before pre-training has commenced
\item \textbf{After pre-training}: Once pre-training is complete, but before post-training (which can involve a number of processes, including fine-tuning to improve performance in specific domains, applying safety features and improving efficiency)
\item \textbf{Before internal deployment}: Before allowing the model to be used internally for research acceleration or other purposes. Internal use could also be approved for specific purposes but not others
\item \textbf{Permission expansion points}: Before granting the model additional access to particular systems, data, or computational resources
\end{enumerate}

\textbf{Safety cases}

We recommend that technical teams within the project be required to make a robust, affirmative safety case to the Board before progressing beyond each of in a list of predetermined thresholds. If they cannot make such a case, either because the system is not sufficiently safe or because it is not yet possible to make an affirmative case with our current level of understanding, then the project should not progress beyond that checkpoint.

A safety case is a structured argument, supported by empirical evidence, that a system is safe to be deployed in a particular environment. They form part of an affirmative safety model which is common in other high-risk industries, where developers are required to demonstrate that risk falls below an acceptable threshold. For example, the Nuclear Regulatory Commission specifies that reactor designs must show that the expected frequency of large-scale core meltdowns is less than 1 in 10,000 years\cite{noauthor_safety_nodate-1}.

Existing regulatory frameworks for AI, such as NIST's AI Risk Management Framework \cite{noauthor_ai_2021}, fall short of assigning specific risk tolerances. The Board should address this gap by setting acceptable risk thresholds for AI models developed by the project.

Thresholds could vary depending on the specific risks the model might pose -- for example, the project might have a lower tolerance for irreversible loss-of-control risk than for the likelihood of power concentration.\footnote{See Affirmative safety: An approach to risk management for high-risk AI \cite{wasil_affirmative_2024} for illustrative risk thresholds.} In principle, there are several ways that technical teams might go about proving acceptably low risk, such as through demonstrable and comprehensive understanding of model internals \cite{doshi-velez_towards_2017}, or through formal proofs of safety \cite{tegmark_provably_2023} that can be mathematically verified.

Safety cases should also include emergency response plans that technical teams will employ in the event that sudden risks emerge after internal deployment. Again, this is a practice employed by the NRC, which requires ``onsite and offsite emergency plans to assure that adequate protective measures can be taken to protect the public in the event of a radiological emergency'' before commercial power plants can be licensed to operate \cite{noauthor_backgrounder_nodate}.

With this said, it is important to acknowledge that gaps in our scientific understanding of AI capabilities mean that building full safety cases for models significantly more advanced than today's is not yet possible \cite{noauthor_safety_nodate}. The project should prioritize resolving the research challenges that would enable comprehensive safety cases, though it remains unclear how long such breakthroughs might take.

\textbf{We believe that both 1) the project should strictly adhere to an affirmative safety standard, as is appropriate given its extremely high-stakes and 2) producing watertight safety cases for very advanced models may not be possible before they can be trained.}

\textbf{This leads to the conclusion that pauses are likely to be necessary}.

Frontier developers have already acknowledged that this may be necessary. An \href{https://www-cdn.anthropic.com/1adf000c8f675958c2ee23805d91aaade1cd4613/responsible-scaling-policy.pdf}{earlier version} of Antrophic's Responsible Scaling Policy states that compliance ``may sometimes require research or technical breakthroughs to give affirmative evidence of a model's safety (which is generally not possible today)'' -- which they say implies a commitment to pause deployment and/ or development if capabilities outpace their ability to produce such evidence.

\pagebreak
\subsection{Board oversight}
\begin{keyrecommendations}
\begin{enumerate}
\item
  Institute a civilian-led board composed of technical experts with the
  competence to thoroughly assess safety cases.
\item
  Establish clear intolerable risk thresholds or ``tripwire
  capabilities'' beyond which deployment or development must be halted
  until effective mitigations are put in place.
\item
  Appoint a point of contact who is the sole intermediary required to
  execute a pause instruction should a member of technical staff request
  one.
\item
  Establish both ``bottom-up'' and ``top-down'' pause protocols, so that
  any member of staff concerned about imminent and extreme risks can
  trigger an initial halt. Require increasingly senior sign-off for
  longer pauses.
\item
  Before resuming training, require technical teams to present a safety
  case to the board, which makes a clearly-evidenced argument for why
  doing so would not pose intolerable risks. Safety cases should include
  emergency response plans to mitigate against extreme risks that emerge
  after internal deployment.
\item
  Require simple majorities for lower-risk, reversible decisions, and
  supermajorities for more consequential and potentially irreversible
  decisions.
\end{enumerate}
\end{keyrecommendations}

High-stakes decisions within the project should be approved by a Board. This Board should be granted legal authority through an act of Congress to permit or disallow various actions by the project. It should also be authorized to impose reporting requirements on technical teams (as described in section 3), as well as granted the same level of access to datacenters, information and personnel as onsite contractors.

\subsubsection{Composition}

\textbf{Despite the strategic significance of AGI for military purposes, we recommend that the Board be civilian-led}, and composed of technical experts with the competence to thoroughly assess affirmative safety cases. There is precedent for civilian-led oversight of strategically significant government projects in the \textbf{Interim Committee}, a body set up within the Manhattan Project in 1945 to advise the President on matters pertaining to nuclear energy. Though the committee was established during wartime, and its first duty was to advise on the use of nuclear weapons against Japan, there was recognition within the small section of the government involved in the Manhattan Project that nuclear energy would have broad civilian applications, and should not be under military control. The same principle should be applied to AGI, which will be a powerful dual-use technology whose potential benefits and dangers are not confined to the military sphere.

The Board for an AGI project should include technical experts with full membership and binding recommendations, ensuring their representation is meaningful rather than tokenistic. This differs from the Interim Committee of the Manhattan Project, which was merely advised by a panel of four physicists who were not full members and whose recommendations were not binding. That panel was partially established to appease scientists concerned about American first-use of nuclear weapons and a potential US-Soviet arms race. Many scientists ultimately disagreed with the panel's consensus on using the bomb against Japan, as evidenced by the Franck Report and Szilard Petition (appeals from Manhattan Project scientists who dissented from the panel's view).

A more instructive model for its Board than the Interim Committee may be the \textbf{\href{https://www.dnfsb.gov/}{Defense Nuclear Facilities Safety Board}}, an independent government agency that oversees the nuclear weapons facilities within the US Department of Energy. The DNFSB is composed of up to five members appointed by the President, who must be confirmed by a majority Senate vote. No more than three seats can be filled by members of the same political party. That members must be ``respected experts in the field of nuclear safety with a demonstrated competence and knowledge relevant to the independent investigative and oversight functions of the Board'' is written into the National Defense Authorization Act (NDAA) for Fiscal Year 1989 \cite{noauthor_national_1988}, which established the DNFSB. The National Academy of Sciences maintains a list of qualified individuals to assist the President in nominating members. The five commissioners that head the \textbf{Nuclear Regulatory Commission} (NRC) are appointed according to a very similar process to the DNFSB \cite{noauthor_42_nodate} in the Energy Reorganization Act of 1974 (though the Act does not specify any particular level of qualification for members).

Appointment of technical experts to the Board of an AGI project could be modelled after the NRC and DNFSB -- though this process will be complicated by the fact that what constitutes expertise in AI (especially its large-scale risks) is far less well-established than in the case of nuclear safety. The likelihood and timing of catastrophic outcomes is hotly debated. For example, the largest survey of published machine learning researchers\cite{grace_survey_nodate} more conservative risk estimates and longer timelines than those working directlty on safety \cite{bensinger_existential_2021}, while AGI lab researchers \cite{noauthor_anthropics_nodate} predict much sooner emergence than academics. The project should be cautious about equating general machine learning publication records with qualifications to assess risks from general-purpose frontier models. Informal polling at NeurIPS (the Conference on Neural Information Processing Systems, the premier annual gathering for machine learning research) showed many researchers weren't familiar with the term ``artificial general intelligence.''\cite{gao_tweet_2024} A survey of 111 AI experts\cite{field_why_2025} found that most hadn't heard of basic catastrophic risk concepts, with familiarity strongly predicting risk perception.

Candidates for Board membership should be assessed according to their specific understanding of and expertise in the potential catastrophic risks of AI stemming from loss-of-control or egregious misuse. There are several existing organizations staffed by researchers with specialized expertise and with independence from AGI labs (many or even most employees of labs may have been drafted into the project itself). The \href{https://www.nist.gov/aisi}{Center for AI Standards and Innovation} (CAISI) is one obvious source of technical expertise. US-based independent AI safety research organizations include \href{https://metr.org/}{Model Evaluation \& Threat Research} (METR), \href{https://www.redwoodresearch.org/}{Redwood Research} and \href{https://far.ai/}{Far AI}. NIST has also partnered with a number of organizations conducting relevant work, that form its \href{https://www.nist.gov/aisi/artificial-intelligence-safety-institute-consortium/aisic-members}{AI Safety Institute Consortium}. Expertise in Test, Evaluation, Verification, and Validation (TEVV) may also be important for assessing safety cases. The project should consider individuals with experience in TEVV specifically in the context of defense technology, such as representatives from Defense Advanced Research Projects Agency (DARPA) or the Test Resource Management Center (TRMC).

\subsubsection{Authorities}

\textbf{The Board's main authorities should be to approve or disapprove a) the training of a large model at a new compute threshold and b) use of an internally deployed model for a series of pre-specified research applications}. \textbf{Its regulatory authority to do so should be established through an act of Congress}. \textbf{As far as possible, the Board's jurisdiction should be limited to determining the \textit{technical safety} of training new models or deploying existing ones in particular research contexts, rather how to use AI capabilities for strategic purposes.}\footnote{We acknowledge that, in practice, there is not a bright line between decisions about the technical safety of a model's development and deployment and its application for strategic purposes. For example, concerns about maintaining competitiveness with adversaries could influence risk tolerance, even if inadvertently.

Models could also be internally deployed for strategically relevant applications, such as forecasting or intelligence. To a degree, the Board may be able to assess the safety of internal deployment for, say, forecasting in isolation, by assessing the model's tendency to hallucinate misleading predictions. But other questions relevant to the wisdom of using AI-generated forecasts are more wide-ranging, such as the potential for forecasts to become self-fulfilling prophecies, or how adversaries might perceive and respond to US forecasting capabilities. Our recommendation is to seek separation between these two domains \textit{as far as possible}, but we acknowledge that a perfect divide may not always be feasible or desirable.

More generally, the concentration-of-power risks that would emerge from a government-led AGI project are beyond the scope of this report but should be given serious consideration. These risks already exist in the status quo\cite{davidson_ai-enabled_2025}, where frontier models are being trained by a small handful of powerful tech companies.}

This is a departure from the precedents set by the Interim Committee and DNFSB. The former played a purely advisory role, while the latter has \textit{some} legally enshrined authorities (the DoE must grant the DNFSB access to nuclear facilities and must respond to its recommendations, but is not obligated to accept them). Perhaps the closest analogy is the NRC's 5-member Commission, which formulates binding policies governing nuclear reactor and materials safety. In practice, the legal authorities or the NRC are delegated among its nearly 3,000 employees, who are responsible for overseeing the country's 54 nuclear power plants. Licensing for individual plants, for example, is usually done at the regional level through one of the NRC's Atomic Safety and Licensing Boards, while the Commission is the final decision-making authority in litigating any disputes or appeals.

In the context of an AGI project, the Board would be overseeing a far more centralized effort and could itself directly arbitrate key decisions. There may be multiple research efforts occurring within the project at any one time (for example, numerous smaller training runs might be undertaken simultaneously for hyperparameter optimization, ablation studies or debugging), but the greatest risks are likely to emerge from training and internally deploying the project's largest, frontier-pushing model -- over which the Board should exercise the most oversight. This may raise concerns about power concentration, which can be partially mitigated by limiting the Board's authority as far as possible to assessing a system's technical safety. It should not extend to geopolitical decisions about how to use AI capabilities for strategic purposes. The Board should be able to mandate a halt in the development or deployment of models, but not be the ultimate arbiter of, for example, if and how to negotiate with adversaries in the aftermath of a pause. This would be the purview of the wider policy establishment and the Geopolitics Division (see section 6).

\textbf{The project could consider requiring simple majorities for lower-risk, reversible decisions, and supermajorities for more consequential and potentially irreversible decisions}. Examples of the former could be approval of new evaluation techniques or changes to reporting requirements, while examples of the latter could be internal deployment of a newly trained model, or resuming model training after a pause.

\subsection{Audit and risk monitoring}

\begin{keyrecommendations}
\begin{enumerate}
\item
  Establish two separate bodies: an internal audit team (to scrutinize
  the alignment techniques and wider risk management practices being
  employed by the project) and a risk monitoring team (to estimate the
  level of risk being posed by the project at any one time).
\item
  Consider designating a Special Inspector General with the specialized
  technical expertise needed to perform internal audit.
\item
  The risk monitoring team should estimate the amount of catastrophic
  risk being posed by the project at each major checkpoint (such as
  pre-training, post-training and internal deployment of a model), as
  well as forecasting risk levels several steps in advance.
\end{enumerate}
\end{keyrecommendations}

In addition to the Board, the project should institute two functions within the project that are responsible for lower-level oversight of its activities: an \textbf{internal audit mechanism} that is responsible for ensuring compliance with the project's safety protocols and scrutinizing the effectiveness of its risk management practices, and a \textbf{risk monitoring team} that estimates the amount of catastrophic risk being posed by the project at any one time.

\subsubsection{Internal audit}

\textbf{The project should implement robust internal audit processes that a) ensure that its safety protocols are followed and b) scrutinize the effectiveness of its risk management processes.}

There is broad expert consensus \cite{schuett_new_2023} on the desirability of internal auditing at frontier AI companies. A government-led project aiming to build AGI should have a similar function. In the US federal government, the primary internal auditing function is performed by an Office of Inspector General (OIG) within each agency. The Inspector General Act of 1978 \cite{noauthor_inspector_nodate} establishes these offices and grants them the statutory authority to access agency records, subpoena information necessary to perform their function and receive and investigate disclosures from whistleblowers (often via dedicated hotlines). Government agencies are also audited by the Government Accountability Office (GAO), which investigates the use of public funds and sets standards for internal audit across the entirety of the federal government.

To some degree, a government-led AGI project could be subject to scrutiny from both these bodies. If the project takes the form of a defense contract with the US government, it could fall under the oversight of the \href{https://www.dodig.mil/}{Department of Defense Office of Inspector General}. Promisingly, Inspector Generals have the broad authority under the IG Act to investigate and report any ``substantial and specific danger to the public health and safety'', which \textit{could} cover catastrophic AI risks.\footnote{This is the same wording that is used in the Whistleblower Protection Act, which, as discussed in section 1.23, may prove problematic when applied to the AI context -- since the nature and timing of risks from powerful AI remains disputed, and they can be difficult to predict ahead of time.} The Act also specifically requires the Secretary of Defense to submit semiannual reports to Congress detailing the findings of the DoD OIG, which must then be made public within 60 days, excluding material that may constitute a threat to national security if released. This could provide a useful oversight mechanism for the project, while still allowing for the redaction of strategically significant information from public reports. The GAO has also historically played a role in assessing the effectiveness of public private partnerships. For example, it produced a report into Covid-19 vaccine development efforts under Operation Warp Speed in 2021.\cite{office_operation_2021}

Oversight from the OIG and GAO could provide a useful foundation for audit of a government-led AGI project. However, stress-testing the risk management practices of such a project will require deep technical expertise that is unlikely to be present in either of these bodies.\footnote{As a detailed report into the application of internal audit to frontier AI development notes, individuals with both AI and audit expertise are rare, due to the relatively recent emergence of AI risk assessment as a practice and its lack of well-established parameters.} Therefore, we recommend that the project designate a specialized team to carry out this function. One option would be to institute a designated OIG to oversee the AGI project. In the past, the government has set up ``Special Inspector Generals'' tailored to particular federal programs. These include the \href{https://www.sigar.mil/}{Special Inspector General for Afghanistan Reconstruction} (SIGAR) and the \href{https://www.oversight.gov/inspectors-general/special-inspector-general-troubled-asset-relief-program}{Special Inspector General for the Troubled Asset Relief Program} (SIGTARP). These perform narrower monitoring of specific government efforts when department-level OIG oversight is not considered sufficient. The pool of expertise that the project can draw from in staffing such a body is similar to that recommended for the Board (see section 3.1). For example, independent safety organizations such as METR have been commissioned by OpenAI and Anthropic \cite{noauthor_update_2023} to evaluate their frontier models, thus playing something of an external audit function (though they have not identified themselves as doing this per se). Risk management personnel from other high-stakes industries with internal or external audit experience could also bring relevant expertise.

\textbf{The core purpose of this team should be to scrutinize the risk management efforts of the project as a whole and identify potential failure modes or gaps in risk coverage}. It could be modelled after Anthropic's Alignment Stress-Testing Team, which describes itself as ``to red-team Anthropic's alignment techniques and evaluations, empirically demonstrating ways in which [they] could fail''\cite{evhub_introducing_2024}. This is especially important in the context of AI development, because superficial risk mitigations that \textit{appear} effective can conceal larger failures. For example, situationally aware models that seem aligned during pre-deployment evaluations may be engaging in deception \cite{park_ai_2023}, sycophancy, \cite{sharma_towards_2025}, or alignment techniques that work well during training but break down when models encounter novel situations they weren't trained on.\cite{lang_distribution_2022} Proposed solutions to these problems include "meta-level" oversight techniques \cite{shlegeris_meta-level_2023} that identify failures in oversight processes themselves, rather than in model behavior.

\subsubsection{Risk monitoring team}

\textbf{The project should dedicate a team to estimating the amount of catastrophic risk being posed by the project at each major checkpoint (such as pre-training, post-training and internal deployment of a model), as well as forecasting risk levels several steps in advance.}

The optimal degree of separation between risk \textit{assessment} and risk \textit{management} within government and other high-stakes environments has been a subject of some debate over the past few decades. A 1983 report on risk assessment within the US federal government from the National Research Council (which focuses mainly on the management of hazardous chemicals) draws attention to the risks of blurring the distinction between these two elements\cite{health_organizational_1983}. Risk management involves \textbf{value judgements} about the ideal tradeoffs between the consequences of particular high-risk activities and interventions to prevent them, whereas the goal of risk assessment is simply to describe the level of risk as accurately as possible. Therefore, lack of a clear separation between the two can distort risk perception and undermine the perceived legitimacy of risk assessments.

The report emphasizes the need for a clear conceptual distinction between these two functions, but stops short of suggesting a full separation between the staff responsible for each, citing the risk of inhibited communication between them. However, it is our view that given the tremendous stakes of AGI development and the potential for geopolitical considerations to distort incentives and risk perception, establishing a team whose sole responsibility is to assess risk posed by the project is appropriate. Decisions that fall under the broader category of risk management (such as which safety research agendas to pursue, whether to train or deploy new models and if and when to initiate international diplomacy) should be the purview of the Board, safety teams, project leadership and the Geopolitics Division. Frequent information sharing between the safety, internal compliance and risk monitoring teams, for example through weekly cross-functional meetings, can ensure adequate knowledge transfer.

\textbf{At designated checkpoints throughout model training and deployment (see section 2.3 for illustrative examples), the risk monitoring team should produce a report that quantifies the approximate probability of each potential adverse outcome that could result from advancing to the subsequent checkpoint}. These could include loss-of-control and the emergence of capabilities enabling egregious misuse. The report should also estimate the total level of a) catastrophic risk and b) existential risk\footnote{There is some debate as to what would constitute an existential vs catastrophic risk. Classical definitions categorize a catastrophic risk as one that ``might have the potential to inflict serious damage to human well-being on a global scale \cite{currie_working_2018}'' and an existential risk as a sub-category of catastrophic risk that ``“threatens the entire future of humanity\cite{bostrom_existential_2013}'', either by causing our extinction or some other irreversible curtailment of potential. The project should establish its own clear definitions, and perhaps a delineation of increasingly severe catastrophic risks based on, for example, expected number of casualties.} posed by continued development or deployment.

As has been a recurring theme in this report, this proposal suffers from the issue that we do not have well-established methods for forecasting the likelihood of AI-caused catastrophes. There is extreme diversity of opinion \cite{rosenberg_roots_2022} on this topic amongst both domain experts and professional forecasters. Calibration may improve as the risk monitoring team gathers more evidence from the development of increasingly powerful models, and near-term quantification of risks emerging from the next stage of development will likely be more accurate than attempts to forecast over very long time horizons. There is also evidence that AIs themselves may soon achieve superhuman forecasting capabilities \cite{noauthor_superhuman_nodate}. The risk monitoring team could make use of internally-developed models for this purpose, subject to approval from the Board.

\subsection{A designated verification project}

\begin{keyrecommendations}
\begin{enumerate}
\item
  Institute a designated project within one of the Department of
  Energy's national labs for the development of assurance methods that
  could be used to verify compliance with treaties on restricting the
  development of powerful AI.
\item
  This project should both accelerate R\&D into novel verification
  technologies (such as hardware-enabled mechanisms) that are capable of
  real-time monitoring \emph{and} explore the application of existing
  verification methods to the AI context.
\item
  Bring in outside expertise to accelerate the development of HEMs,
  through a defense contract with NVIDIA.
\item
  Consider pursuing collaboration between this verification project and
  similar efforts in other countries, such as the UK's Safeguarded AI
  Project.
\end{enumerate}
\end{keyrecommendations}

As emphasized from the beginning of this paper, a central priority of the project should be \textit{maintaining the option} of pivoting away from building AGI as quickly as possible, and towards international coordination around delaying its development. If both the US and China, for example, become convinced that continuing to develop powerful AI will be imminently risky, measures that slow down or change the default trajectory of development may be considered desirable by all sides. However, international agreements on AI will be difficult to design or sustain if compliance cannot be verified between geopolitic adversaries.

The US government should therefore institute a \textbf{designated project for the development of assurance methods} that could be used to verify compliance with treaties on restricting the development of powerful AI. This project could aim to 1) accelerate R\&D into novel verification technologies such as \cite{petrie_technical_2025} that are capable of real-time monitoring, and 2) explore the application of \cite{wasil_verification_2024} including on-site inspections and remote sensing to the AI context.

\subsubsection{Implementation}

A verification team could be placed ``outside'' the core AGI project and require less stringent security clearances than those ``inside'' it, since verification technologies lack the strategic sensitivity of powerful AI models, and can be developed independently of them. This allows for easier integration of the effort within existing government bodies, less need for stringent security, and the potential for international collaboration (as discussed below).

\textbf{A natural home for a verification project could be within one of the Department of Energy's national labs}. The Sandia National Laboratory already has ``Remote Sensing and Verification\cite{noauthor_global_nodate}'' as one of its priority research areas under the umbrella of Global Security, making it a possible candidate for hosting this initiative. The project could draw on existing expertise from within the lab to research, for example, how energy monitoring could be leveraged to detect unauthorized training runs, or how remote sensing techniques could identify undeclared datacenters.

\subsubsection{Hardware enabled mechanisms}

While traditional assurance methods should be explored for the AI case, there are important disanalogies between AI and previous strategically significant technologies which may necessitate novel technological approaches to ensure robust verification. For example, the Treaty on the Non-Proliferation of Nuclear Weapons (NPT) relies partly on on-site inspections that measure uranium enrichment levels at nuclear facilities\cite{noauthor_verification_2016}. These inspections can differentiate between permitted civilian energy production (requiring lower enrichment) and prohibited weapons development (requiring highly enriched uranium). AI verification presents a different challenge -- inspectors at an AI datacenter would find no physical traces distinguishing between permitted and prohibited computations. Additional factors such as the potential for discontinuous improvements in AI systems, the phenomenon of emergent capabilities \cite{wei_emergent_2022}, and the unpredictable relationship between performance and training compute \cite{heim_training_2024} call for verification technologies that monitor computations in real time -- and ideally prevent, rather than simply detect, rule violations.

The term ``Hardware Enabled Mechanisms'' (HEMs) refers to a suite of possible technologies that could embed regulatory controls directly onto AI chips. The most comprehensive exploration of HEMs comes from the RAND Corporation \cite{kulp_hardware-enabled_2024}, which introduces two varieties -- \textbf{offline licensing}, which would require chips to possess a specialized license in order to operate, and \textbf{fixed set}, which would impose network restrictions on GPUs to prevent their aggregation into large clusters. The report primarily proposes HEMs as a means of augmenting US export controls, but also discusses their use as a potential means of verifying multilateral agreements. It concludes that HEM technologies are possible in principle but that they may take several years to develop and deploy. A more concrete proposal \href{https://yoshuabengio.org/wp-content/uploads/2024/08/FlexHEG-Memo_August-2024.pdf}{introduces} \textbf{Flexible Hardware Enabled Mechanisms (FlexHEGs)}, programmable devices that would allow for the enforcement of a wide variety of multilaterally agreed rules, and refuse to operate when asked to perform disallowed functions. An initial report makes the case that compelling FlexHEG prototypes could be developed with around 18 months of concerted R\&D effort.

\textbf{The government should bring in outside expertise to accelerate the development of HEMs. A defense contract with NVIDIA, which the government could compel the company to prioritize using Defense Production Act powers, could direct some of the world's top talent and resources in cutting-edge chip production to the problem of designing hardware-based verification mechanisms.}

\subsubsection{International cooperation}

The government could also consider pursuing collaboration between this verification project and similar efforts in other countries. The nature of verification technology incentivizes collaboration rather than competition, since all countries stand to benefit from its standardized adoption. Relevant expertise exists in multiple countries (the UK's \href{https://www.aria.org.uk/opportunity-spaces/mathematics-for-safe-ai/safeguarded-ai/}{Safeguarded AI project}, the US's \href{https://www.galois.com/}{Galios}, France's \href{https://compcert.org/}{CompCert}), and pooling of international talent and resources could accelerate progress. Historically, adversaries have collaborated on the development of verification mechanisms even as they competed to build strategically significant technologies. For example, the US and the Soviet Union participated in the 1988 \href{https://nonproliferation.org/lab-to-lab-joint-verification-experiment/}{Joint Verification Experiment} (JVE), which developed and tested seismic monitoring techniques to verify compliance with the 1974 Threshold Test Ban Treaty. The \href{https://www.ctbto.org/our-work/international-monitoring-system}{International Monitoring System} (IMS), a global network of over 330 monitoring facilities designed to detect nuclear explosions, was developed by the Comprehensive Nuclear-Test-Ban Treaty Organization, which has 187 member states.

\subsection{Intelligence and Scenario Planning Division}

\begin{keyrecommendations}
\begin{enumerate}
\item
  Establish a separate body responsible for gathering intelligence
  relevant to adversary progress towards AGI, and scenario planning for
  geopolitical crises emerging from AGI development
\item
  Ensure separation between this body and the teams responsible for
  building AGI, to avoid conflicts of interest.
\item
  Develop detailed plans for scenarios including (but not limited to) a
  US pause, adversary threats of military action over AGI development,
  China achieving AGI, model weight theft and internal power grabs.
\end{enumerate}
\end{keyrecommendations}

\textbf{The government should designate a body responsible for a) tracking the progress of US adversaries, particularly China, towards AGI and b) planning for a set of predetermined scenarios in which it will be necessary to engage in high-stakes diplomacy with foreign powers to prevent dangerous escalation and/ or loss of control over powerful AI.}

This could take the form of a designated office within the Department of Defense, similar to the Office of Net Assessment which was responsible for conducting projections of military, technological, and political trends relevant to national security, or the Defense Threat Reduction Agency (DTRA), which detects and prevents WMD threats to the United States.

This body should be strictly separate from the teams actually overseeing the development of AGI, to avoid conflicts of interest, for example, incentivise exaggerated reports of adversary progress to justify faster development.

\subsubsection{Intelligence on adversary progress towards AGI}

Several Manhattan Project scientists later came to believe that the less-than-cautious development of the atomic bomb was driven largely by inaccurate assessments of German and Soviet progress toward nuclear weapons \cite{belfield_why_nodate}. Key participants even came to regret their work. The project had a higher risk tolerance than it might have otherwise -- accepting, for example, some risk (if small) of igniting the atmosphere during the Trinity Test.

It is critical that a government-led AGI project avoid this same mistake. Falsely believing that it is in an arms race, or miscalculating the closeness of such a race, could compel its leaders to take risks that increase the probability of catastrophic outcomes. For example, there is evidence that the extent of Chinese AI progress has been historically overestimated by Western observers. The release of DeepSeek's impressive R1 model in January 2025 sparked panic about a so-called ``Sputnik moment'' \cite{hawkins_who_2025} for AI, and cause a sudden plummet in US tech stock prices -- but later analysis suggested that its performance could have been predicted by naive extrapolation of scaling laws \cite{amodei_deepseek_nodate}, and whether its headline training cost was accurately reported is up for debate \cite{roeloffs_what_nodate}. Several analysts made the case that DeepSeek's success strengthened the need to curtail China's AI efforts using interventions such as export controls, rather than compelling the US to accelerate its own.

Therefore, the project should dedicate significant intelligence resources to assessing the true state of Chinese AI progress.\footnote{The project should maintain smaller-scale efforts to monitor frontier AI progress in other countries, but the primary focus should remain on China, which, at the time of writing, is the US's only realistic adversarial competitor in the AI space.} Some of this work can be done using publicly available information, and the government may be able to accelerate it through funding relevant think tanks. For example, the DC-based Center for Security and Emerging Technology has published research on US and allied dominance of the semiconductor supply chain, quantified Chinese contributions to high-impact AI research and studied US retention of Chinese STEM graduates \cite{zwetsloot_trends_nodate}. Others will require expertise and capabilities unique to the intelligence community. The Intelligence and Scenario Planning Division could track Chinese AI efforts by, for example, monitoring energy consumed by data centers through grid data, tracking the efficacy of export controls and the countervailing impacts of chip smuggling, and maintaining lists of the most important Chinese entities and individual researchers in the AI ecosystem.

\subsubsection{Crisis scenario planning}

The development of powerful AI could trigger geopolitical crises that require urgent de-escalation. A key responsibility of the Intelligence and Scenario Planning Division should be developing detailed scenario plans that can be employed in such crises. These plans should be operationalized to a similar level of detail to, for example, the Single Integrated Operational Plan (SIOP), which served as the United States' general plan for nuclear war between 1961 and 2003.

Detailed plans should be made for the following scenarios:

\begin{enumerate}
\item \textbf{US pause}: The US-led project has paused its frontier training due to imminent catastrophic risk. The concern has been deemed credible by the Board, and it does not appear that it will be possible to proceed safely for some time. In this situation, the US should disclose to China that it has paused, ideally provide verifiable evidence of the capabilities of concern, and initiate efforts towards a non-proliferation treaty.

\item \textbf{Threat of war}: The US government should be prepared for any nuclear-armed adversary to threaten kinetic action up to and including nuclear strikes if it becomes convinced that the project is close to achieving a decisive strategic advantage with powerful AI.\footnote{Adversaries may engage in less extreme tactics in order to sabotage a US AGI project before resorting to overt threats of conflict, such as espionage or cyberattacks. A \href{https://www.nationalsecurity.ai/chapter/deterrence-with-mutual-assured-ai-malfunction-maim}{recent report} refers to such tactics as Mutual Assured AI Malfunction or MAIM attacks, and makes the case that, by default, we should expect nations to climb this escalation ladder should its early attempts at sabotage fail.} The Geopolitics Division should prepare detailed plans for how to handle such scenarios.

\item \textbf{Chinese AGI}: The Geopolitics Division should prepare for a scenario in which there is compelling evidence that China has achieved, or is about to achieve, AGI before the US.

\item \textbf{Model weight theft}: The project must have emergency plans to respond to the theft of model weights by a well-resourced state actor.\footnote{The project should invest considerably into securing model weights in line to prevent such a scenario from occuring in the first place. Weights cannot be unreleased once they have been stolen or proliferated. Nonetheless, the project should plan for crisis diplomacy with the actors in possession of the weights, should security measures fail.}

\item \textbf{Internal coup}: Extremely powerful AI systems could enable unprecedented concentrations of power among the small number of individuals overseeing them. The project should prepare for a scenario in which a group of actors within it attempt such a power grab. This may require emergency shutdowns or military intervention.
\end{enumerate}

\subsection{A plan for automated research}

\begin{keyrecommendations}
\begin{enumerate}
\item
 Develop a clear plan for the use of future automated researchers that are capable of accelerated AI research. 

\item
  Draw clear red lines regarding how much authority can be ceded to AI systems

\item
  Make a list of safety agendas and protocols to be pursued with the aid of automated researchers.

\end{enumerate}
\end{keyrecommendations}

\textbf{Safety teams within the project should develop a clear plan for the use of future AI systems that are capable of radically accelerating and eventually automating its own research. This plan should be subject to approval by the Board}. It should include:

\begin{enumerate}
\item Clear red lines regarding how much authority can be ceded to AI systems and the acceptable limits of AI-enabled acceleration
\item A list of safety agendas and experimental protocols to be pursued with the aid of automated researchers
\end{enumerate}

AI systems may eventually surpass humans at the task of AI R\&D itself, likely leading to radical acceleration of AI development. A recent study \cite{noauthor_evaluating_2024} found that AI models are already competitive with human AI researchers over 2-hour time horizons, though humans maintain an edge on tasks completed over longer periods. The safety plans released by each frontier lab track the degree to which their systems can accelerate the pace of AI R\&D as a potentially dangerous capability\cite{noauthor_common_2025}.

At present, AI companies do not have detailed, publicly auditable plans as to how they will use the automated AI researchers they will eventually develop, though the incentives to use them for the purposes of accelerating their own models' capabilities will be extremely strong. In a recent essay, Anthropic CEO Dario Amodei hinted at the possibility of attaining a competitive advantage through allowing models to improve themselves: ``because AI systems can eventually help make even smarter AI systems, a temporary lead could be parlayed into a durable advantage''\cite{amodei_deepseek_nodate}.

Such a strategy is clearly fraught with risk. The pace of automated AI R\&D could accelerate dramatically as AI systems can copy themselves, creating what Nobel Prize and Turing Award winner Geoffrey Hinton calls a digital ``hive mind'' \cite{morris_ai_nodate} of artificial researchers that both outpace and outnumber their human overseers. Effective human-in-the-loop monitoring will become increasingly challenging. For example, if automated AI R\&D enabled a 30x speed-up in the pace of development, it could require 30x or more man-hours of review and oversight than previously to keep pace with the AI. The project should place a hard limit on the speed-up that can be enabled by internally deployed models to ensure adequate human oversight.

Automated researchers can be deployed for safety purposes as well as capabilities enhancement -- research agendas such as OpenAI's now abandoned Superalignment \cite{noauthor_introducing_2023} strategy are illustrative. Significant effort should be invested in forecasting how automated researchers might contribute to safety. The project should compile comprehensive research agendas and experimental protocols in preparation for the arrival of systems that can meaningfully accelerate safety research.

\section{Conclusion}

We believe that implementing some or all of the features described in the context of a government-led AGI project could reduce catastrophic risk stemming from loss-of-control or egregious misuse of powerful systems. However, a recurring theme throughout this report has been the scientific immaturity of current methods used to evaluate and control AI models. Scientific breakthroughs are needed before an AGI project will be able to develop smarter-than-human AI systems in accordance with the protocols outlined in this report -- which will require writing watertight safety cases for models much more powerful today, developing a comprehensive method for detecting dangerous capabilities, and ultimately maintaining control over AIs more capable than ourselves.

Until such breakthroughs occur, adherence with these protocols would effectively delay the development of AGI. This is a possibility that we believe that project should accept and be actively preparing for.

\hfill
\hfill
\hfill

\subsubsection*{Acknowledgements}
This project benefited greatly from the mentorship of Elliot Jones, as well as thoughtful comments from:

Morgan Simpson, Liam Patell, Eli Lifland, Connor A. Stewart Hunter, Mauricio Baker, Peter Wildeford, Nick Marsh, Bruna Avellar, Neha Suresh, and William Fowler.

This paper does not necessarily represent the views of acknowledged individuals.

\bibliographystyle{IEEEtranN}
\bibliography{references}

\end{document}